\def\booltwo{0} 
\def\booldraft{0} 

\if\booltwo1
\documentclass[10pt]{article}%
\usepackage[letterpaper, total={8in, 9in}]{geometry}
\else
\documentclass[11pt]{article}%
\usepackage[letterpaper, total={6.5in, 9in}]{geometry}
\fi

\usepackage{amsmath}
\usepackage{graphicx}

\usepackage{amsfonts}
\usepackage{amssymb}%
\usepackage{mathtools}

\usepackage{calligra} 
\DeclareMathAlphabet{\mathcalligra}{T1}{calligra}{m}{n}
\DeclareFontShape{T1}{calligra}{m}{n}{<->s*[2.2]callig15}{}
\newcommand{\R}{R}

\usepackage{currfile} 

\usepackage{tikz}
\usetikzlibrary{calc}
 
\usepackage{caption}
\usepackage{subcaption}

\usepackage[normalem]{ulem}

\usepackage[english]{babel}

\newcommand{\D}[0]{\mbox{d}}

\newcommand{\GeneralDerivative}[4]
{ \frac{ {#4}^{#3} {#1} }{ {#4} {#2}^{#3} } }

\newcommand{\pderiv}[2]{ \GeneralDerivative{}{#1}{#2}{\partial} }

\newcommand{\commute}[2]{ \left[ {#1} \, , {#2} \right] }

\newcommand{\ket}[1]{ \left| {#1} \right \rangle }
\newcommand{\bra}[1]{ \left \langle {#1} \right| }

\newcommand{\Outer}[2]{ \left| {#1} \middle\rangle \middle\langle  {#2} \right| }

\newcommand{\subsc}[1]{ \mbox{\tiny{#1}} } 
\newcommand{\subsn}[1]{ \scalebox{0.5}{#1} } 

\title{Causality in the Fermi Problem and the Magnus expansion}

\date{\today{}}

\begin{document}

\if\booltwo1
\twocolumn[
\fi

\begin{center}

{\Large{}Causality in the Fermi Problem and the Magnus expansion}

\bigskip

\centering
J.\ S.\ Ben-Benjamin%

\medskip

\centering%
\textit{Institute for Quantum Science and Engineering, Texas A\&M University, Texas, USA.}

%
%
%

\medskip

\bigskip

{\centering{}\bfseries{}Abstract}

\medskip

\parbox{5in}
{

In 1932,
Fermi presented a two-atom model
for determining whether quantum mechanics
is consistent with causality,
and concluded that
indeed it is.
In the late 1960's,
Shirokov and others
found that Fermi's approximations
may not have been sound,
and when corrected,
Fermi's model shows non-causal behavior.
We show that if
instead of time-dependent perturbation theory,
the Magnus expansion is used to approximate the time-evolution operator,
causality does follow.
} 
\end{center}
\medskip

\if\booltwo1
]
\fi

\if\booldraft1
\noindent%
The current file is: \currfilename.\\
\fi

\section{Introduction}

\begin{figure}[b!]
\centering
\begin{tikzpicture}[thick, scale=1.5]
\def\dist{3};
\def\radi{0.5};
\def\sml{0.3};
\def\lg{0.8};
\filldraw [gray] (0    ,0) circle (\radi cm);
\filldraw [gray] (\dist,0) circle (\radi cm);
\node at (0    ,\radi+\sml) {$A_{\subsc L}$};
\node at (\dist,\radi+\sml) {$A_{\subsc R}$};
\draw[thick,->] (-\lg,-\radi-\sml)--(\dist+\lg,-\radi-\sml);
\node at (\dist+\lg+\sml,-\radi-\sml) {$z$};
\node at (0    ,-\radi-2*\sml) {$z_{\subsc L}$};
\node at (\dist,-\radi-2*\sml) {$z_{\subsc R}$};
\draw[thick] (0    ,-\radi-\sml-0.4*\sml)--(0    ,-\radi-\sml+0.4*\sml);
\draw[thick] (\dist,-\radi-\sml-0.4*\sml)--(\dist,-\radi-\sml+0.4*\sml);
\node at (\dist/2,\sml*2/3) {$\R$};
\draw[dotted,thick] (0,0) -- (\dist,0);
\node at (\dist/2,-\radi-\sml-\lg) {$\ket{\psi(t=0)}=\ket{b,a,0}$};
\end{tikzpicture}
\caption{
Initial conditions of the Fermi model.
Two atoms are separated by a distance $\R$:
The one on the left,
$A_{\subsc L}$ at position $z_{\subsc L}$,
is initially in the ground state,
and the one on the right,
$A_{\subsc R}$ at position $z_{\subsc R}$,
is initially in the excited state;
no photons are present.
The initial state is
$\ket{\psi(t=0)}=\ket{b,a,0}$,
where
$\ket0$ is the field vacuum state,
and
$\ket b$ and $\ket a$
are the ground and excited atomic states, respectively.
} 
\label{fig:01-1}
\end{figure}
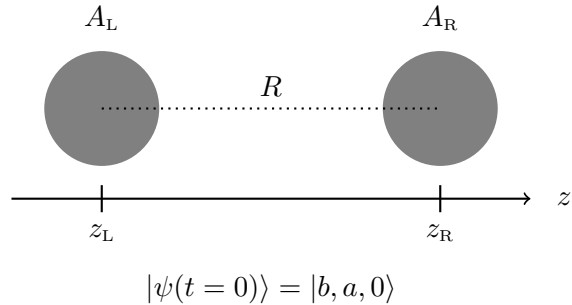

In his 1932 review article
on the quantum theory of radiation,
Fermi saught to show that
quantum theory gives causal results
\cite{fermi}.
Fermi
devised
the following model:
At time $t=0$,
a pair of stationary two-level atoms,
$A_{\subsc R}$ and $A_{\subsc L}$,
a distance $\R$ apart,
where
atom $A_{\subsc R}$ is initially in the excited state
and
atom $A_{\subsc L}$ is initially in the ground  state,
and no photons are present
(See Fig.\ \ref{fig:01-1}).
Immediately after time $t=0$,
the probability that
atom $A_{\subsc R}$
is in the ground state
becomes non-zero.
The question Fermi asked is:
When does the probability amplitude,
$\mathcal A$,
that
$A_{\subsc L}$
is excited
become non-zero?

The model involves
two spacetime events:
Atom $A_{\subsc R}$
starts decaying at
$(t_{\subsc R},z_{\subsc R})$
and
atom $A_{\subsc L}$
starts becoming excited at
$(t_{\subsc L},z_{\subsc L})$;
quantum mechanics is consistent with causality if
the probability amplitude $\mathcal A$ is zero for
$|t_{\subsc L}-t_{\subsc R}| < |z_{\subsc L}-z_{\subsc R}|/c$.
Using his model,
Fermi found that
indeed,
quantum theory
is consistent with causality.

Sometime later,
Shirokov showed that
Fermi made some simplifying assumptions
in his calculation,
without which,
his model
yields non-causal predictions
\cite{shirokov}.
Specifically,
Shirokov found that
the ground state atom
could become excited instantaneously.
These results
have been widely discussed,
and
views
generally
belong to
one of the following:
(a)
Quantum theory should give causal answers,
including in the Fermi model
\cite{fermi,heitler1,heitler2,heitler3,ScullyZubairy},
(b)
the
Fermi model
leads to absurd results
because it
is non-physical
\cite{shirokov,heger,hegerPRL},
(c)
the
Fermi model is not physically-realizable,
but
if one modifies it,
causal predictions 
could be obtained
\cite{ferretti,shirokov-78},
and
(d)
quantum theory's prediction of
an instantaneous effect
is physically correct.

The present paper is
in
the first camp.
In Sec.\ \ref{sec:magnusFermi}
we
show that causality is restored
in the Fermi model if
the state is evolved using
the Magnus expansion
of the time-evolution operator.
In Sec.\ \ref{sec:nonCaus}
we discuss the rotating wave approximation
as it relates to causality.
To illustrate the Magnus expansion method
for a simple case,
in the appendix
we present an exactly solvable probem.
\section{The Fermi problem and the Magnus expansion}
\label{sec:magnusFermi}

To find the probability amplitude,
\begin{align}
%
\mathcal A
=
\langle
	{a,b,0}
	|
	{\hat U(t,0)}
	|
	{b,a,0}
\rangle
,
\end{align}
that the state
$\ket{b,a,0}$
at time $t=0$
evolves into the state
$\ket{a,b,0}$
at some later time $t$,
we use the Magnus expansion method.
As we explain in
the appendix,
to second order,
the Magnus expansion
gives that
the time-evolution operator
is approximately the sum of three terms,
\begin{align}
%
\hat U = \hat 1 + \hat M_1 + \hat M_2
.
\label{eq:09-37}
\end{align}
%
Explicitly,
Eq.\ \eqref{eq:09-37} is
\cite{magnus2},
\begin{align}
\hat U(t,0)
=
\hat 1
+
\int_0^t \frac{\D t'}{i\hbar} \hat V(t')
+
\int_0^t \frac{\D t'}{i\hbar}
\int_0^{t'} \frac{\D t''}{i\hbar}
\commute
	{ \hat V(t' ) }
	{ \hat V(t'') }
,
\label{eq:01-35}
\end{align}
where the interaction Hamiltonian
is
$\hat V = \hat V_{\subsc L} + \hat V_{\subsc R}$,
with
\begin{align}
\hat V_j(t)
=
g_j
\left( \Big.
	\hat \sigma
	e^{-i\omega t}
	+
	\mbox{H.a.}
\right)
\sum_\nu
\sqrt\nu
\left( \Big.
	\hat a_\nu
	e^{-i \nu(ct \pm z_j)/c}
	+
	\mbox{H.a.}
\right)
,
\label{eq:08-91}
\end{align}
and where $j$ is `L' or `R'
(standing for the left and right atoms respectively),
and
the $(\pm)$ sign corresponds to
left-moving and right-moving waves.
In Eq.\ \eqref{eq:08-91},
$g_j=\wp_j\sqrt{\hbar/\epsilon_{\subsn0}L}$,
and
the operators
$\hat\sigma^\dagger=\Outer ab$
and
$\hat\sigma=\Outer ba$
are the atomic raising and lowering operators.
The field creation and annihilation operators of frequency $\nu$,
$\hat a_\nu^\dagger$
and
$\hat a_\nu$,
interact locally
and therefore are
evaluated
at the atomic position $z_j$.

Since the Fermi model
involves two atomic state changes,
two contributions of
the interaction Hamiltonian are required.
Therefore
only the third term in Eq.\ \eqref{eq:01-35},
$\hat M_2$,
contributes to the amplitude $\mathcal A$,
%
\begin{align}
%
\mathcal A
=
\int_0^t \D t' \,
\int_0^{t'} \D t'' \,
C(z',t';z'',t'')
,
\end{align}
where $C(z',t';z'',t'')$ is the matrix element of the
commutator
$
\bra{a,b,0}
[
\hat V(t'),
\hat V(t'')
]
\ket{b,a,0}
$,
which involves
two-spacetime events.
The matrix element $C$ is
\begin{align}
C(z',t';z'',t'')
&=
\pm
\frac
	{ c\hbar \wp_{\subsc L} \wp_{\subsc R} }
	{ i\pi\epsilon_{\subsn0} }
\pderiv {\R}{}
\Big\{
	e^{ i\omega_{\subsc L}t' - i\omega_{\subsc R}t'' }
	\delta
	\left( \Big.
		t''-t'
		\pm
		\R/c
	\right)
\nonumber
\\&-
	e^{ i\omega_{\subsc L}t'' - i\omega_{\subsc R}t' }
	\delta
	\left( \Big.
		t'-t''
		\pm
		\R/c
	\right)
\Big\}
,
\label{eq:06-33}
\end{align}
where
$\R=z_{\subsc R}-z_{\subsc L}$.
We therefore have that
the probability amplitude $\mathcal A$
for the Fermi model is
\begin{align}
\mathcal A
&=
\pm
\frac
	{ c\hbar \wp_{\subsc L} \wp_{\subsc R} }
	{ i\pi\epsilon_{\subsn0} }
\pderiv {\R}{}
\int_0^t \!\! \D t'
\int_0^{t'} \!\! \D t'' \,
\Big\{
	e^{ i\omega_{\subsc L}t' - i\omega_{\subsc R}t'' }
	\delta
	\left( \Big.
		t''-t'
		\pm
		\R/c
	\right)
\nonumber
\\&-
	e^{ i\omega_{\subsc L}t'' - i\omega_{\subsc R}t' }
	\delta
	\left( \Big.
		t'-t''
		\pm
		\R/c
	\right)
\Big\}
.
\label{eq:06-34}
\end{align}
The spacetime $\delta$-functions
enforce that $\mathcal A$ remain exactly zero
until at least a time $\R/c$ has passed,
thus giving a causal prediction.
Thus,
the spacetime $\delta$-functions
are crucial.
These
appear
because
the electric field operator is special
and
obeys Maxwell equations,
which are causal.

Since $\R$ is positive
and since
the integration limits restrict $t' \ge t''$,
we find that
the first $\delta$-function in
Eq.\ \eqref{eq:06-34}
could have a zero argument only for left-moving waves
($+$ sign),
and that the second $\delta$-function
could have zero argument only for right-moving waves
($-$ sign).
We note that
even though
the first  $\delta$-function gives
the left-moving wave contribution
and
the second $\delta$-function gives
the right-moving wave contribution,
it is \emph{not}
the case that
the first  consists of only      co-rotating terms
and that
the second consists of only counter-rotating terms.
Actually,
each $\delta$-function requires both
co-rotating and counter-rotating terms,
as we discuss in Sec.\ \ref{sec:nonCaus}.

These two processes
are not distinguishable,
and therefore add in amplitude.
The left-propagating term is
\begin{align}
\mathcal A_+
=
\frac
	{ c\hbar \wp_{\subsc L} \wp_{\subsc R} }
	{ i\pi\epsilon_{\subsn0} }
\pderiv {\R}{}
\Theta(t-\R/c)
e^{ i\omega_{\subsc R} \R/c}
\frac
	{
		e^{ i( \omega_{\subsc L} -\omega_{\subsc R} )t }
		-
		e^{ i( \omega_{\subsc L} -\omega_{\subsc R} )\R/c }
	}
	{\omega_{\subsc L} -\omega_{\subsc R} }
,
\label{eq:07-35}
\end{align}
and the right-propagating term is
\begin{align}
\mathcal A_-
=
\frac
	{ c\hbar \wp_{\subsc L} \wp_{\subsc R} }
	{ i\pi\epsilon_{\subsn0} }
\pderiv {\R}{}
\Theta(t-\R/c)
e^{-i\omega_{\subsc L} \R/c}
\frac
	{
		e^{ i( \omega_{\subsc L} -\omega_{\subsc R} )t }
		-
		e^{ i( \omega_{\subsc L} -\omega_{\subsc R} )\R/c }
	}
	{\omega_{\subsc L} -\omega_{\subsc R} }
.
\label{eq:07-36}
\end{align}
Both Amplitudes \eqref{eq:07-35} and \eqref{eq:07-36}
exhibit causality
because they are exactly zero for $t<\R/c$.
The total probability amplitude
for the Fermi problem is
therefore
$\mathcal A = \mathcal A_+ + \mathcal A_-$,
\begin{align}
\mathcal A
&=
-\frac
	{ c\hbar \wp_{\subsc L} \wp_{\subsc R} }
	{ \pi\epsilon_{\subsn0} }
\pderiv {\R}{}
\Theta(t-\R/c)
\left( \Big.
	e^{ i\omega_{\subsc R} \R/c}
	+
	e^{-i\omega_{\subsc L} \R/c}
\right)
\nonumber
\\&\times
\frac
	{
		e^{ i( \omega_{\subsc L} -\omega_{\subsc R} )t }
		-
		e^{ i( \omega_{\subsc L} -\omega_{\subsc R} )\R/c }
	}
	{\omega_{\subsc L} -\omega_{\subsc R} }
,
\end{align}
which is causal.

%
%
%
%
%
%
%
%
%
%
%
%
%
%

%
%
%
\section{The Rotating Wave Approximation}
\label{sec:nonCaus}

There are two contributions to the probability amplitude
for the Fermi problem:
(a)
Atom $A_{\subsc R}$
could transition to the ground state,
emitting a photon
which after a time $\R/c$ excites
atom $A_{\subsc L}$,
and
(b)
atom $A_{\subsc L}$
becomes excited and emits a photon,
which a time $\R/c$ later
is absorbed by
atom $A_{\subsc R}$,
which transitions to the ground state.
The second
is the counter-rotating process,
and not only does it contribute,
it is in fact essential for causality.

We know that the quantized radiation Hamiltonian
contains both co-rotating and counter-rotating terms,
and that both have an essential role
\cite{ScullyZubairy};
neglecting the counter-rotating terms
(an approximation called
``the rotating-wave approximation,''
or RWA)
is only appropriate
in specific situations
and only for calculating certain quantities
\cite{RWAbad1,RWAbad2,RWAbad4,RWAfreq,RWAbad3}.
For example,
even in the case in which
the RWA is considered to be most appropriate
(in the near-resonant two-level atom case),
the counter-rotating terms have
a significant contribution to the frequency shift
of individual atoms
\cite{RWAfreq}.
This shows
that
ignoring the counter-rotating terms
leads to
non-causal
results.

In the interaction Hamiltonian,
Eq.\ \eqref{eq:08-91},
the terms that go as
$\hat\sigma\hat a^\dagger$
and
$\hat\sigma^\dagger\hat a$,
are called co-rotating.
It also contains
counter-rotating terms;
those go like
$\hat\sigma^\dagger\hat a^\dagger$
and
$\hat\sigma\hat a$,
and do not (individually) conserve energy.
The RWA
amounts to neglecting the latter two
terms,
which would lead
to the interaction Hamiltonian
$\hat V = \hat V_{\subsc L} + \hat V_{\subsc R}$,
where
\begin{align}
\hat V_j(t)
\overset{ \subsc{RWA} }{\longrightarrow}
g_j
\sum_\nu
\sqrt\nu
\left( \Big.
	\hat\sigma_j \hat a_\nu^\dagger
	e^{-i[\omega t - \nu(ct \pm z_j)/c]}
	+
	\mbox{H.a.}
\right)
.
\label{eq:05-31}
\end{align}
One can show that using
this purely co-rotating
interaction Hamiltonian
yields non-causal predictions. 
Specifically,
the integrand in the Magnus expansion
consists of functions other than $\delta$-functions.
As discussed in Sec.\ \ref{sec:magnusFermi},
to be causal,
the integrand must consist of only $\delta$-functions.
The details will be published in a future paper.

%
%
%
\section{Conclusion}

The Fermi model has been widely used for discussing the issue of causality in quantum theory.
While Fermi showed that the model implies causality,
Shirokov and others have shown that it does not.
We have shown that
if one uses the Magnus expansion,
one obtains causality in a straightforward way.
Also,
we have found that the the rotating wave approximation leads to non-causal results.
That the Magnus expanstion method
gives causal results
in the Fermi model
may imply that
causality is related to operator-ordering,
as suggested in Ref.\ \cite{schleich};
this is because
the Magnus expansion uses a different operator-ordeing
than TDPT.

\section{Acknowledgements}

We would like to thank Professor
M.\ O.\ Scully
for insightful discussions,
the Robert A.\ Welch Foundation (Grant No.\ A-1261),
the Office of Naval Research (Award No.\ N00014-16-1-3054),
the Air Force Office of Scientific Research (FA9550-18-1-0141),
and
the King Abdulaziz City for Science and Technology (KACST) grant
for their the support.

\appendix

\section*{Appendix: An exactly solvable model illustrating the Magnus method}
\label{sec:exact}

We now present an exactly solvable model
and use it to compare
the Magnus expansion
to
time-dependent perturbation theory (TDPT).
While causality is not an issue here,
this model allows us to explicitly contrast
the Magnus expansion method and TDPT.

Consider the Hamiltonian
for a quantized single-mode radiation field
due to a classical current,
$ 
\hat H
=
\hbar\omega \hat a^\dagger \hat a
+
g( \hat a^\dagger + \hat a)
$,
which leads to the interaction potential
\begin{align}
\hat V(t)
=
g
(
	\hat a^\dagger
	e^{i\omega t}
	+
	\hat a
	e^{-i\omega t}
)
.
\label{eq:01-37}
\end{align}
The exact time-evolution operator
is 
\begin{align}
\hat U_{\subsc{Exact}}(t,0)
&=
\exp
\left[ \Big.
	\frac g{\hbar\omega}
	\left\{ \Big.
		\hat a
		\left( \Big. e^{-i\omega t} - 1 \right)
		-
		\hat a^\dagger
		\left( \Big. e^{ i\omega t} - 1 \right)
	\right\}
\right]
.
\label{eq:01-40b}
\end{align}
%
%

We now
contrast
the exact time-evolution operator
and
the approximations obtained by TDPT
and by the Magnus expansion method.

\bigskip
\noindent%
\textbf{Magnus expansion.}
Since every unitary operator
is the exponential of some anti-Hermitian operator
\cite{Stone},
the time-evolution operator,
which is unitary,
may be written as
$\hat U = \exp[\hat M]$.
Following Magnus
\cite{magnus1,magnus2},
we write
the exponent as
$\hat M = \hat M_1 + \hat M_2 + \cdots$.
In the Magnus expansion method,
we truncate this sum
and then expand the exponential in a power series,
resulting in Eqs.\ \eqref{eq:09-37} and \eqref{eq:01-35}.

For the potential in Eq.\ \eqref{eq:01-37},
the
terms $\hat M_1$ and $\hat M_2$ are
[see Eq.\ \eqref{eq:01-35}]
\begin{align}
\hat M_1
=
\frac g{\hbar\omega}
\left\{ \Big.
	\hat a
	\left( \Big. e^{-i\omega t} - 1 \right)
	-
	\hat a^\dagger
	\left( \Big. e^{ i\omega t} - 1 \right)
\right\}
,
\end{align}
and
\begin{align}
\hat M_2
&=
\frac{ig^2}{(\hbar\omega)^2}
\left( \Big.
	\omega t
	-
	\sin (\omega t)
\right)
.
\end{align}
Since $\hat M_2$
is proportional to the indentity operator,
the exponential form of $\hat U$ is
$
\exp[\hat M_1 + \hat M_2]
=
\exp[\hat M_2]
\exp[\hat M_1]
$,
where we have truncated the exponent
to the first two terms
$\hat M \simeq \hat M_1 + \hat M_2$,
but the resulting operator is still unitary.
Explicitly,
\begin{align}
\hat U
\simeq
\exp
\left[ \Big.
	\frac{ig^2}{(\hbar\omega)^2}
	\left( \Big.
		\omega t
		-
		\sin (\omega t)
	\right)
\right]
\exp
\left[ \Big.
	\frac g{\hbar\omega}
	\left\{ \Big.
		\hat a
		\left( \Big. e^{-i\omega t} - 1 \right)
		-
		\hat a^\dagger
		\left( \Big. e^{ i\omega t} - 1 \right)
	\right\}
\right]
.
\label{eq:17-40}
\end{align}

Eq.\ \eqref{eq:17-40} is almost the same as the exact 
time-evolution operator
$\hat U_{\subsc{Exact}}$ in
Eq.\ \eqref{eq:01-40b}:
The two expressions become equal
if we approximate $\sin(\omega t) \simeq \omega t$,
that is,
$\hat M_2 \simeq 0$.
However,
should we want to approximate Eq.\ \eqref{eq:17-40} further,
we would keep the overall scalar factor,
$\exp[\hat M_2]$,
and expand the exponential of $\hat M_1$ in a power series.
This approach gives
\begin{align}
\hat U_{ \subsc{Mag} }
&=
\exp
\left[ \Big.
	\frac{ig^2}{\hbar^2\omega^2}
	\left( \Big.
		\omega t
		-
		\sin (\omega t)
	\right)
\right]
\nonumber
\\&\times
\left\{ \Big.
	\hat 1
	+
	\frac g{\hbar\omega}
	\left[ \Big.
		\left( \Big. e^{-i\omega t} - 1 \right)
		\hat a
		-
		\left( \Big. e^{ i\omega t} - 1 \right)
		\hat a^\dagger
	\right]
\right\}
,
\label{eq:01-48a}
\end{align}
which is close to unitary:
When calculating
$\hat U_{ \subsc{Mag} }^\dagger \hat U_{ \subsc{Mag} }$,
the leading term is the identity $\hat 1$,
the term first-order in $g$ is zero,
and the second-order term oscillates like
$\sin^2(\omega t/2)$ with time
and decreases with the square of the photon energy.
Explicitly
\begin{align}
\hat U_{ \subsc{Mag} }^\dagger \hat U_{ \subsc{Mag} }
=
\hat 1
+
\left( \Big.
	\frac g{\hbar\omega}
\right)^2
\sin^2 \left( \Big. \frac{\omega t}2 \right)
\left[ \Big.
	e^{ i\omega t/2} \hat a^\dagger
	+
	e^{-i\omega t/2} \hat a
\right]^2
,
\end{align}
which is approximately unitary
for large photon frequency $\omega$.

\bigskip
\noindent%
\textbf{Time-dependent perturbation theory.}
To second-order,
TDPT gives
\begin{align}
&
\hat U_{ \subsc{TDPT} }
=
\hat 1
+
\frac g{\hbar\omega}
\left[ \Big.
	\left( \Big. e^{ i\omega t} - 1 \right)
	\hat a^\dagger
	-
	\left( \Big. e^{-i\omega t} - 1 \right)
	\hat a
\right]
\nonumber
\\&+
\frac{g^2}{\hbar^2\omega^2}
\Big[
	\left( \Big.
  		i\omega t
		+
		e^{-i\omega t} - 1
	\right)
	\hat a \hat a^\dagger
	-
	\left( \Big.
  		i\omega t
		-
		e^{ i\omega t} + 1
	\right)
	\hat a^\dagger \hat a
\nonumber
\\&-
	\left( \Big.
  		\frac{ e^{-2i\omega t} - 1 }2
		+
		e^{-i\omega t} - 1
	\right)
	\hat a \hat a
	+
	\left( \Big.
  		\frac{ e^{ 2i\omega t} - 1 }2
		-
		e^{ i\omega t} + 1
	\right)
	\hat a^\dagger \hat a^\dagger
\Big]
.
\label{eq:01-45}
\end{align}
From Eq.\ \eqref{eq:01-45},
we see that the TDPT
time-evolution operator increases linearly in time,
that is,
$\hat U_{ \subsc{TDPT} }(t,0) \propto t$.
Therefore,
TDPT yields predictions for the total probability
that diverge as $t^2$.
In contrast,
the Magnus expansion approximation
for the time-evolution operator,
$\hat U_{ \subsc{Mag} }$,
does not diverge
in this way.
This,
of course,
does not show that the Magnus expansion method
is always better than TDPT.

%
%
%
%
%

%
%
%
%
%
%
%

%
%
%
%

%
%
%
%
%
%
%

%
%
%
%
%
%
%
%
%
%

\begin{thebibliography}{9}
%

\bibitem{fermi}
{
E.\ Fermi,
``Quantum Theory of Radiation,''
Rev.\ Mod.\ Phys.\ 4 87 (1932).
}


%
\bibitem{shirokov}
{
M.\ I.\ Shirokov,
``The Velocity of Electromagnetic Retardation in Quantum Electrodynamics,''
Sov.\ J.\ Nulc.\ Phys.\ 4 774 (1967).
}

\bibitem{heitler1}
{
W.\ Heitler, S.\ T.\ Ma,
``Quantum theory of radiation damping for discrete states,''
Proc.\ R.\ Ir.\ Acad.\ 52 123 (1949)%
.
}

\bibitem{heitler2}
{
J.\ Hamilton,
``Damping Theory and the Propagation of Radiation,''
Proc.\ Phys.\ Soc.\ A 62 12 (1949)%
.
}

\bibitem{heitler3}
{
W.\ Heitler,
\textit{The Quantum Theory of Radiation},
Oxford (1964)%
.
}

\bibitem{ScullyZubairy}
{
M.\ O.\ Scully, M.\ S.\ Zubairy,
\textit{Quantum Optics},
Cambridge Univ.\ Press, New York (1997)%
.
}


%




\bibitem{heger}
{
G.\ C.\ Hegerfeldt,
``Remark on causality and particle localization,''
Phys.\ Rev.\ D 10 3320 (1974).
}

\bibitem{hegerPRL}
{
G.\ C.\ Hegerfeldt,
``Causality Problems for Fermi's Two-Atom System,''
Phys.\ Rev.\ Lett.\ 72 5 596 (1994).
}

\bibitem{ferretti}
{
B.\ Ferretti,
``Propagation of Signals and Particles,''
in \textit{Old and New Problems in Elementary Particles},
Acad.\ Press, New York (1968)%
.
}

\bibitem{shirokov-78}
{
M.\ I.\ Shirokov,
``Signal velocity in quantum electrodynamics,''
Sov.\ Phys.\ Usp.\ 21 345 (1978)%
.
}


\bibitem{magnus1}
{
W.\ Magnus,
``On the exponential solution of differential equations for a linear operator,''
Commun.\ Pure Appl.\ Math.\ VII 649 (1954).
}

\bibitem{magnus2}
{
S.\ Blanes, F.\ Casas, J.\ A.\ Oteo, J.\ Ros,
``The Magnus expansion and some of its applications,''
Phys.\ Rep.\ 470 151 (2009).
}






\bibitem{Stone}
{
M.\ H.\ Stone,
``On one-parameter unitary groups in Hilbert Space,''
Ann.\ Math.\ 33 (3) 643 (1932).
}

\bibitem{RWAbad1}
{
L.\ Mandel, D.\ Meltzer,
``Theory of Time-Resolved Photoelectric Detection of Light,''
Phys.\ Rev.\ 188, 198 (1969)%
.
}

\bibitem{RWAbad2}
{
G.\ S.\ Agarwal,
``Rotating-wave approximation and spontaneous emission,''
Phys.\ Rev.\ A 4 1778 (1971)%
.
}

\bibitem{RWAbad4}
{
G.\ S.\ Agarwal,
\textit{Quantum Optics},
Springer, Berlin (1974).
}

\bibitem{RWAfreq}
{
P.\ L.\ Knight, L.\ Allen,
``Rotating-wave approximation in coherent interactions,''
Phys.\ Rev.\ A 7 368 (1973).
}

\bibitem{RWAbad3}
{
H,\ Zheng, S.-Y.\ Zhu, M.\ S.\ Zubairy,
``Quantum Zeno and anti-Zeno effects: without the rotating-wave approximation,''
Phys.\ Rev.\ Lett.\ 101 200404 (2008)%
.
}


%
%
%

\bibitem{schleich}
{
L.\ I.\ Plimak, S.\ T.\ Stenholm, W.\ P.\ Schleich,
``Operator ordering and causality,''
Phys.\ Scr.\ T147 (2012).
}


%
%
%


%
\end{thebibliography}
\end{document}